\documentstyle[12pt]{article}

\font\twlgot =eufm10 scaled \magstep1 \font\egtgot =eufm8
\font\sevgot =eufm7 \font\twlmsb =msbm10 scaled \magstep1
\font\egtmsb =msbm8 \font\sevmsb =msbm7

\newfam\gotfam
\def\pgot{\fam\gotfam\twlgot}
\textfont\gotfam\twlgot \scriptfont\gotfam\egtgot
\scriptscriptfont\gotfam\sevgot
\def\got{\protect\pgot}
\newfam\msbfam
\textfont\msbfam\twlmsb \scriptfont\msbfam\egtmsb
\scriptscriptfont\msbfam\sevmsb
\def\Bbb{\protect\pBbb}
\def\pBbb{\relax\ifmmode\expandafter\Bb\else\typeout{You cann't use
Bbb in text mode}\fi}
\def\Bb #1{{\fam\msbfam\relax#1}}

\newcommand{\gd}{{\got d}}

\newcommand{\ccG}{{\got g}}

\def\thebibliography#1{\section*{References}\list
  {[\arabic{enumi}]}{\settowidth\labelwidth{#1}\leftmargin\labelwidth
    \advance\leftmargin\labelsep
    \usecounter{enumi}}
    \def\newblock{\hskip .11em plus .33em minus .07em}
    \sloppy\clubpenalty4000\widowpenalty4000
    \sfcode`\.=1000\relax}

\def\op#1{\mathop{\fam0 #1}\limits}

\newcommand{\id}{{\rm Id\,}}

\newcommand{\beq}{\begin{equation}}
\newcommand{\eeq}{\end{equation}}
\newcommand{\ben}{\begin{eqnarray}}
\newcommand{\een}{\end{eqnarray}}
\newcommand{\be}{\begin{eqnarray*}}
\newcommand{\ee}{\end{eqnarray*}}
\newcommand{\bea}{\begin{eqalph}}
\newcommand{\eea}{\end{eqalph}}

\newcommand{\cA}{{\cal A}}

\newcommand{\cP}{{\cal P}}
\newcommand{\cD}{{\cal D}}
\newcommand{\cR}{{\cal R}}
\newcommand{\cL}{{\cal L}}

\newcommand{\cE}{{\cal E}}
\newcommand{\cH}{{\cal H}}

\newcommand{\cS}{{\cal S}}
\newcommand{\cO}{{\cal O}}
\newcommand{\cN}{{\cal N}}
\newcommand{\cG}{{\cal G}}
\newcommand{\cQ}{{\cal Q}}
\newcommand{\bL}{{\bf L}}

\newcommand{\bb}{{\bf 1}}

\newcommand{\al}{\alpha}

\newcommand{\dl}{\delta}
\newcommand{\la}{\lambda}
\newcommand{\La}{\Lambda}
\newcommand{\f}{\phi}
\newcommand{\om}{\omega}

\newcommand{\m}{\mu}

\newcommand{\g}{\gamma}
\newcommand{\G}{\Gamma}
\newcommand{\th}{\theta}

\newcommand{\vt}{\vartheta}
\newcommand{\vf}{\varphi}
\newcommand{\up}{\upsilon}
\newcommand{\lng}{\langle}
\newcommand{\rng}{\rangle}

\newcommand{\im}{{\rm Im\,}}

\newcommand{\si}{\sigma}
\newcommand{\Si}{\Sigma}
\newcommand{\w}{\wedge}
\newcommand{\wt}{\widetilde}
\newcommand{\wh}{\widehat}
\newcommand{\ol}{\overline}
\newcommand{\dr}{\partial}
\newcommand{\ar}{\op\longrightarrow}
\newcommand{\ot}{\otimes}
\newcommand{\ap}{\approx}
\newcommand{\ve}{\varepsilon}
\newcommand{\intr}{{\rm Int\,}}
\newcommand{\re}{{\rm Re\, }}

\newcounter{eqalph}
\newcounter{equationa}
\newcounter{remark}
\newcounter{example}
\newcounter{theorem}
\newcounter{proposition}
\newcounter{lemma}
\newcounter{corollary}
\newcounter{definition}
\def\theremark{\arabic{remark}}
\def\thetheorem{\arabic{theorem}}

\def\thedefinition{\arabic{definition}}

\newenvironment{eqalph}{\stepcounter{equation}
\setcounter{equationa}{\value{equation}} \setcounter{equation}{0}

\begin{eqnarray}}{\end{eqnarray}
\setcounter{equation}{\value{equationa}}}

\newcommand{\mar}[1]{}

\begin{document}
\hbox{}

{\parindent=0pt

{\large \bf Axiomatic quantum field theory. Jet formalism}
\bigskip

{\sc G. Sardanashvily}

{\sl Department of Theoretical Physics, Moscow State University,
117234 Moscow, Russia}

\bigskip
\bigskip

\begin{small}

{\bf Abstract.} Jet formalism provides the adequate mathematical
formulation of classical field theory reviewed in
hep-th/0612182v1. A formulation of QFT compatible with this
classical one is discussed. We are based on the fact that an
algebra of Euclidean quantum fields is graded commutative, and
there are homomorphisms of the graded commutative algebra of
classical fields to this algebra. As a result, any variational
symmetry of a classical Lagrangian yields the identities which
Euclidean Green functions of quantum fields satisfy.

\end{small}

 }

\bigskip
\bigskip

Jet manifold formalism provides the adequate mathematical
formulation of classical field theory, called axiomatic classical
field theory (henceforth ACFT) (see \cite{axiom} for a survey).
Bearing in mind quantization, we consider a graded
$C^\infty(X)$-module of even and odd classical fields on a smooth
manifold $X=\Bbb R^n$, $n\geq 2$, coordinated by $(x^\la)$. In
ACFT, these fields are represented by sections $s$ of a graded
vector bundle $Y=Y_0\oplus Y_1\to X$ coordinated by $(x^\la,
y^a)$. Finite order jet manifolds $J^rY$, $r=1,\ldots,$ of $Y\to
X$ are also vector bundles over $X$ coordinated by $(x^\la,y^a,
y^a_\La)$, $|\La|=k\leq r$, where $\La=(\la_1,\ldots,\La_k)$
denote symmetric multi-indices. Sections of $Y$ and the jet
bundles $J^rY\to X$ generate a graded commutative
$C^\infty(X)$-algebra $\cP^0$ of polynomials of $(y^a, y^a_\La)$.
The differential graded algebra $\cP^*$ of differential forms (the
Chevalley--Eilenberg differential calculus) over $\cP^0$ is split
into the Grassmann-graded variational bicomplex, describing
Lagrangian theory of fields $s$ (see Appendix A).

Our goal here is a formulation of QFT compatible with this
classical one. We follow the concept of algebraic quantum theory
where a quantum system is characterized by a topological
involutive algebra $A$ and positive continuous forms $f$ on $A$,
i.e., $f(a^*a)\geq 0$, $a\in A$. If $A$ is a Banach algebra
admitting an approximate identity (in particular, a
$C^*$-algebra), the Gelfand--Naimark--Segal (henceforth GNS)
representation theorem associates to any positive continuous form
on $A$ a cyclic representation of $A$ by bounded (continuous)
operators in a Hilbert space \cite{dixm}. Note that there are
different extensions of the GNS representation theorem
\cite{book05}.

In QFT, one deals with unnormed topological involutive algebras
represented by unbounded operators (see Appendix B). The GNS
representation theorem  is generalized to these algebras as
follows \cite{hor,schmu}. Let $A$ be a unital topological
involutive algebra and $f$ a positive continuous form on $A$ such
that $f(\bb)=1$ (i.e., $f$ is a state). There exists a strongly
cyclic Hermitian representation $(\pi_f,\th_f)$ of $A$ such that
$f(a)=\lng \pi(a)\th_f|\th_f\rng$, $a\in A$. The representation
$\pi(A)$ of $A$ is an $Op^*$- algebra.

There are two main approaches to axiomatic formulation of QFT. In
the framework of the first approach, called local QFT, one
associates to a certain class of subsets of a Minkowski space a
net of von Neumann, $C^*$- or $Op^*$-algebras which obey some
axioms \cite{araki,buch,haag,halv,hor}. Its inductive limit is
called either a global algebra (in the case of von Neumann
algebras) or a quasilocal algebra (for a net of $C^*$-algebras).
This construction is extended to non-Minkowski spaces, e.g.,
globally hyperbolic spacetimes \cite{brun,brun2,ruzz}.

We follow a different formulation of axiomatic QFT where quantum
field algebras are tensor algebras. Let $Q$ be a nuclear space
(see Appendix C). Let us consider the direct limit
\mar{x2}\beq
A_Q=\wh\ot Q =\Bbb C \oplus Q \oplus Q\wh\ot Q\oplus \cdots
Q^{\wh\ot n}\oplus\cdots \label{x2}
\eeq
of the vector spaces $\wh\ot^{\leq n} Q =\Bbb C \oplus Q \oplus
Q\wh\ot Q\cdots \oplus Q^{\wh\ot n}$, where $\wh\ot$ is the
topological tensor product with respect to the Grothendieck's
topology (which coincides with the $\ve$-topology on the tensor
product of nuclear spaces \cite{piet}). The space (\ref{x2}) is
provided with the inductive limit topology, the finest topology
such that the morphisms $\wh\ot^{\leq n} Q\to \wh\ot Q$ are
continuous and, moreover, are imbeddings \cite{trev}. A convex
subset $V$ of $\wh\ot Q$ is a neighborhood of the origin in this
topology iff $V\cap \wh\ot^{\leq n} Q$ is so in $\wh\ot^{\leq n}
Q$. Furthermore, one can show that $A_Q$ (\ref{x2}) is a unital
nuclear barreled LF-algebra \cite{belang} (see Appendix B). The
LF-property implies that a linear form $f$ on $A_Q$ is continuous
iff the restriction of $f$ to each $\wh\ot^{\leq n} Q$ is so
\cite{trev}. If a continuous conjugation $*$ is defined on $Q$,
the algebra $A_Q$ is involutive with respect to the operation
\mar{x3}\beq
*(q_1\ot\cdots \ot q_n)=q_n^*\ot\cdots q_1^* \label{x3}
\eeq
on $Q^{\ot n}$ extended by continuity and linearity to $Q^{\wh\ot
n}$. One can show that $A_Q$ is a $b^*$-algebra as follows. Since
$Q$ is a nuclear space, there is a family $\|.\|_k$, $k\in\Bbb
N_+$, of continuous norms on $Q$. Let $Q_k$ denote the completion
of $Q$ with respect to the norm $\|.\|_k$. Then the tensor algebra
$\ot Q_k$ is a $C^*$-algebra and  $A_Q$ (\ref{x2}) is the
projective limit of these $C^*$-algebras with respect to morphisms
$\ot Q_{k+1}\to \ot Q_k$ \cite{igu}.

Since $A_Q$ is a nuclear barreled $b$-algebra, one can apply to it
the following variant of the GNS representation theorem. Let $A$
be a unital nuclear barreled $b^*$-algebra and $f$ a positive form
on $A$. There exists a unique cyclic representation $\pi_f$ of $A$
in a Hilbert space by operators on a common invariant domain $D$
\cite{igu}. This domain can be topologized to conform a rigged
Hilbert space such that all the operators representing $A$ are
continuous on $D$.

In axiomatic QFT, one usually choose $Q$ the Schwartz space of
functions of rapid decrease (see Appendix D). For the sake of
simplicity, we here restrict our consideration to real scalar
fields. One associates to them the Borchers algebra
\mar{qm801}\beq
A=\Bbb R\oplus RS^4\oplus RS^8\oplus\cdots, \label{qm801}
\eeq
where $RS^{4k}$ is the nuclear space of smooth real functions of
rapid decrease on $\Bbb R^{4k}$ \cite{borch,hor}. It is the real
subspace of the space $S(\Bbb R^{4k})$ of smooth complex functions
of rapid decrease on $\Bbb R^{4k}$. Its topological dual is the
space $S'(\Bbb R^{4k})$ of tempered distributions (generalized
functions). Since the subset $\op\ot^kS(\Bbb R^4)$ is dense in
$S(\Bbb R^{4k})$, we henceforth identify $A$ with the tensor
algebra $A_{RS^4}$ (\ref{x2}). Then any continuous positive form
on the Borchers algebra $A$ (\ref{qm801}) is represented by a
collection of tempered distributions $\{W_k\in S'(\Bbb R^{4k})\}$
such that
\mar{qm810}\beq
f(\psi_k)= \int
W_k(x_1,\ldots,x_k)\psi_k(x_1,\ldots,x_k)d^4x_1\cdots d^4x_k,
\qquad \psi_k\in RS^{4k}. \label{qm810}
\eeq
In many cases, the $k$-point distributions $W_k$, $k>2$, are
expressed into the two-point ones $W_2$ due to the Wick theorem
relations
\be
W_k(x_1,\ldots,x_q) =\sum W_2(x_{i_1},x_{i_2}) \cdots
W_2(x_{i_{n-1}},x_{i_k}),
\ee
where the sum runs through all partitions of the set $1,\ldots,k$
in ordered pairs $(i_1<i_2),\ldots(i_{k-1}<i_k)$.

For instance, the states of scalar quantum fields in the Minkowski
space (see Appendix E for the case of free scalar fields) are
described by the Wightman functions $W_n\subset S'(\Bbb R^{4k})$
in the Minkowski space which obey the Garding--Wightman axioms of
axiomatic field theory \cite{bogol,sim74,wigh,zin}. Let us mention
the Poincar\'e covariance axiom, the spectrum condition and the
locality condition. In particular, the Poincar\'e covariance
condition implies the translation invariance and the Lorentz
covariance of Wightman functions. Due to the translation
invariance of Wightman functions $W_k$, there exist tempered
distributions $w_k\in S'(\Bbb R^{4k-4})$, also called Wightman
functions, such that
\mar{1278}\beq
W_k(x_1,\ldots,x_k)= w_k(x_1-x_2,\ldots,x_{k-1}-x_k). \label{1278}
\eeq
Note that Lorentz covariant tempered distributions for one
argument only are well described \cite{bogol,zin07}.

In order to modify Wightman's theory, one studies different
classes of distributions which Wightman functions belong to
\cite{solov,thom}. To involve odd fields, one considers
superdistributions as continuous mappings of a certain space of
superfunctions to a nuclear graded commutative algebra
\cite{naga}.

A problem is that there are still no interacting models of the
Wightman axioms. In QFT, quantum fields created at some instant
and annihilated at another one are described by complete Green
functions.  They are given by the chronological functionals
\mar{1260,030}\ben
&&f^c(\psi_k)= \int
W_k^c(x_1,\ldots,x_k)\psi_k(x_1,\ldots,x_k)d^4x_1\cdots d^4x_k,
\qquad \psi_k\in RS^{4k}, \label{1260}\\
&& W^c_k(x_1,\ldots,x_k)= \op\sum_{(i_1\ldots
i_k)}\th(x^0_{i_1}-x^0_{i_2})
\cdots\th(x^0_{i_{k-1}}-x^0_{i_n})W_k(x_1,\ldots,x_k), \label{030}
\een
where $W_k\in S'(\Bbb R^{4k})$ are tempered distributions, $\th$
is the Heaviside function, and the sum runs through all
permutations $(i_1\ldots i_k)$ of the tuple of numbers
$1,\ldots,k$ \cite{bog2}. A problem is that the functionals
$W^c_k$ (\ref{030}) need not be tempered distributions (see
Appendix D). For instance, $W^c_1\in S'(\Bbb R)$ iff $W_1\in
S'(\Bbb R_\infty)$, where $\Bbb R_\infty$ is the compactification
of $\Bbb R$ by means of the point $\{+\infty\}=\{-\infty\}$
\cite{bogol}. Moreover, the chronological forms are not positive.
Therefore, they do not provide states of the Borchers algebra
$A_{RS^4}$ in general.

At the same time, the chronological forms (\ref{030}) come from
the Wick rotation of Euclidean states of the Borchers algebra
\cite{sard91,ccr,conf} (see Appendix F). As is well known, the
Wick rotation enables one to compute the Feynman diagrams of
perturbative QFT by means of Euclidean propagators. Let us suppose
that it is not a technical trick, but quantum fields in an
interaction zone are really Euclidean. It should be emphasized
that the above mentioned Euclidean states differ from the
well-known Schwinger functions in the Osterwalder--Shraded
Euclidean QFT \cite{bogol,ost,schlin,sim74,zin}. The Schwinger
functions are the Laplace transform of Wightman functions, but not
chronological forms (see Appendix G). Note that the Euclidean
counterpart of time ordered correlation functions is also
considered in the Euclidean quantum field theory, but not by means
of the Wick rotation \cite{guer}. Usually, the Wick rotation in
scalar field theory is studied. There is a problem of describing
the Wick rotation on a curved space-time \cite{jaffe} and in
spinor geometry \cite{mck}. To solve this problem, a complex
space-time can be called into play \cite{esp}.

Since the chronological forms (\ref{030}) are symmetric, the
Euclidean states of the tensor algebra $A_{RS^4}$ can be obtained
as states of the corresponding commutative tensor algebra
$B_{RS^4}$ \cite{sard91,ccr}. Therefore, let $\Phi$ be a nuclear
space and $B_\Phi$ a commutative tensor algebra of $\Phi$.
Provided with the direct sum topology, $B_\Phi$ becomes a
topological involutive algebra. It coincides with the enveloping
algebra of the Lie algebra of the additive Lie group $T(\Phi)$ of
translations in $\Phi$. Therefore, one can obtain the states of
the algebra $B_\Phi$ by constructing cyclic strongly continuous
unitary representations of the nuclear Abelian group $T(\Phi)$
(see Appendix H). Such a representation is characterized by a
continuous positive-definite generating function $Z$ on $\Phi$. By
virtue of the Bochner theorem \cite{bochn,gelf64}, this function
is the Fourier transform
\mar{031}\beq
Z(\phi)=\int \exp[i \langle\phi,w\rangle]d\mu(w) \label{031}
\eeq
of a positive measure $\mu$ of total mass 1 on the topological
dual $\Phi'$ of $\Phi$. Then the above mentioned representation
$\pi$ of $T(\Phi)$ can be given by the operators
\mar{qq2}\beq
\wh\phi u(w)=\exp[i\langle \f,w\rangle]u(w)  \label{qq2}
\eeq
in the Hilbert space $L_{\Bbb C}^2(\Phi',\m)$ of the equivalence
classes of square $\m$-integrable complex functions $u(w)$ on
$\Phi'$. The cyclic vector $\th$ of this representation is the
$\m$-equivalence class $\th\ap_\m 1$ of the constant function
$u(w)=1$.

Conversely, every positive measure $\m$ of total mass 1 on the
topological dual $\Phi'$ of $\Phi$ defines the cyclic strongly
continuous unitary representation (\ref{qq2}) of the group $T(Q)$.
One can show that distinct generating functions $Z$ and $Z'$
characterize equivalent representations $T_Z$ and $T_{Z'}$
(\ref{qq2}) of $T(\Phi)$ in the Hilbert spaces $L^2_{\Bbb
C}(\Phi',\m)$ and $L^2_{\Bbb C}(\Phi',\m')$  iff they are the
Fourier transform of equivalent measures on $\Phi'$.

If the function $\alpha\to Z(\alpha\phi)$ on $\Bbb R$ is analytic
at 0 for each $\phi\in \Phi$, a state $f$ of $B_\Phi$ is given by
the expression
\mar{w0}\beq
f_k(\phi_1\cdots\phi_k)=i^{-k}\frac{\dr}{\dr \al^1}
\cdots\frac{\dr}{\dr\alpha^k}Z(\alpha^i\phi_i)|_{\alpha^i=0}=\int\langle
\phi_1,w\rangle\cdots\langle \phi_k,w \rangle d\mu(w). \label{w0}
\eeq
Then one can think of $Z$ (\ref{031}) as being a generating
functional of complete Euclidean Green functions $f_k$ (\ref{w0}).

For instance, free Euclidean fields are described by Gaussian
states. Their generating functionals are of the form
\mar{a10}\beq
Z(\phi)=\exp(-\frac12M(\f,\f)), \label{a10}
\eeq
where $M(\f,\f)$ is a positive-definite Hermitian bilinear form on
$\Phi$ continuous in each variable. In this case, the forms $f_k$
(\ref{w0}) obey the Wick theorem relations where $f_1=0$ and
$f_2(\f,\f')=M(\f,\f')$. The generating function (\ref{a10}) is
the Fourier transform of some Gaussian measure on $\Phi'$. In
particular, let $\Phi=RS^4$ and $f$ be a Gaussian state of
$B_{RS^4}$ such that the covariance form $M$ is represented by a
distribution $M(x,x')\in S'(\Bbb R^8)$ which is the Green function
of some positive-definite elliptic operator
$L_xM(x,x')=\dl_{x'}(x)$. Then the Gaussian state $f$ describes
Euclidean fields with $M(x,x')$ playing the role of their
propagator.  For instance,
\be
L_x=-\Delta_x +m^2, \qquad
M(x,x')=\int(p^2+m^2)^{-1}\exp(-ip(x-x'))d_4x,
\ee
where $p^2$ is the Euclidean scalar.

A problem is that a measure $\m$ in the generating functional $Z$
(\ref{031}) fail to be written in an explicit form. The familiar
expression
\be
\m=\exp(-\int L(\f)d^4x)\op\prod_x[d\f(x)]
\ee
used in perturbative QFT fails to be a true measure. Note that
there is no (translationally-invariant) Lebesgue measure on
infinite-dimensional vector space as a rule (see \cite{versh} for
an example of such a measure). Here, we are not concerned with
different formulations of functional integrals in QFT
\cite{cart,feld,glimm,john,mick,schmit}, but follow perturbative
Euclidean QFT. This is phrased in terms of symbolic functional
integrals and provide Euclidean Green functions in the Feynman
diagram technique.

Let us consider a Lagrangian system of even and odd fields on
$X=\Bbb R^n$ which is described by the DGA $\cP^*$ with the basis
$\{y^a\}$ (see Appendix A). Let $L\in \cP^{0,n}$ be its Lagrangian
which is assumed to be nondegenerate. If an original Lagrangian is
degenerate, one follows the BV prequantization procedure in order
to obtain a nondegenerate gauge-fixing BRST extended Lagrangian,
depending on original fields and ghosts
\cite{ward06,bat,fust,gom}. We suppose that $L$ is a Lagrangian of
Euclidean fields on $\Bbb R^n$. Let us quantize this Lagrangian
system in the framework of perturbative QFT. Since the generating
functional in perturbative QFT depends on the action functional
one usually replaces horizontal densities, depending on jets, with
local functionals evaluated for the jet prolongations of sections
of $Y\to X$ of compact support \cite{ala,barn,bran97,mccloud}.
Note that such functionals, in turn, define differential forms on
functional spaces \cite{castr3,ferr}. In a different way, we are
based on the fact that an algebra of Euclidean quantum fields is
graded commutative, and there are homomorphisms of the graded
commutative algebra $\cP^0$ of classical fields to this algebra
\cite{ward06,ward06a}.

Let $\cQ$ be the graded complex vector space whose basis is the
basis $\{y^a\}$ for the DGA $\cP^*$.  Let us consider the tensor
product
\mar{z40}\beq
\Phi=\cQ\ot S'(\Bbb R^n) \label{z40}
\eeq
of the graded vector space $\cQ$ and the space $S'(\Bbb R^n)$ of
tempered distributions on $\Bbb R^n$. One can think of elements of
$\Phi$ (\ref{z40}) as being $\cQ$-valued distributions on $\Bbb
R^n$. Let $T(\Bbb R^n)\subset S'(\Bbb R^n)$ be a subspace of
functions $\exp\{ipx'\}$, $p\in \Bbb R_n$, which are generalized
eigenvectors of translations in $\Bbb R^n$ acting on $S(\Bbb
R^n)$. We denote $\f^a_p=y^a\ot\exp\{ipx'\}$. Then any element
$\f$ of $\Phi$ can be written in the form
\mar{z42}\beq
\f(x')=y^a\ot\f_a(x')=\int \f_a(p)\f_p^a d_np, \label{z42}
\eeq
where $\f_a(p)\in S'(\Bbb R_n)$ are the Fourier transforms of
$\f_a(-x')$. For instance, there are the $\cQ$-valued
distributions
\mar{z32,43}\ben
&& \f^a_x(x')=\int \f^a_p e^{-ipx}d_np=y^a\ot \dl(x-x'),
\label{z32}\\
&& \f^a_{x\La}(x')=\int (-i)^k p_{\la_1}\cdots p_{\la_k}\f^a_p
e^{-ipx}d_np. \label{z43}
\een

In the framework of perturbative Euclidean QFT, we associate to a
nondegenerate Lagrangian system $(\cP^*,L)$ the graded commutative
tensor algebra $B_\Phi$ generated by elements of the graded vector
space $\Phi$ (\ref{z40}) and the following state $\lng.\rng$ of
$B_\Phi$. For any $x\in X$, there is a homomorphism
\mar{z45}\beq
\g_x: s_{a_1\ldots a_r}^{\La_1\ldots\La_r} y^{a_1}_{\La_1}\cdots
y^{a_r}_{\La_r} \mapsto s_{a_1\ldots a_r}^{\La_1\ldots\La_r}(x)
\f^{a_1}_{x\La_1}\cdots \f_{x\La_r}^{a_r}, \qquad s_{a_1\ldots
a_r}^{\La_1\ldots\La_r}\in C^\infty(X), \label{z45}
\eeq
of the algebra $\cP^0$ of classical fields to the algebra $B_\Phi$
which sends the generating elements $y^a_\La\in \cP^0$ to the
elements $\f^a_{x\La}\in B_\Phi$, and replaces coefficient
functions $s$ of elements of $\cP^0$ with their values $s(x)$ at a
point $x$. Then the above mentioned state $\lng.\rng$ of $B_\Phi$
is defined by symbolic functional integrals
\mar{z31,',47}\ben
&& \lng \f_1\cdots \f_k\rng=\frac{1}{\cN}\int \f_1\cdots \f_k
\exp\{-\int \cL(\f^a_p)d^nx\}\op\prod_p
[d\phi_p^a], \label{z31}\\
&& \cN=\int \exp\{-\int \cL(\f^a_p)d^nx\}\op\prod_p
[d\phi_p^a], \label{z31'}\\
&& \cL(\f^a_p)=\cL(\f^a_{x\La})=\cL(x,\g_x(y^a_\La)), \label{z47}
\een
where $\f_i$ and $\g_x(y^a_\La)=\f^a_{x\La}$ are given by the
formulas (\ref{z42}) and (\ref{z43}), respectively.  The forms
(\ref{z31}) are expressed both into the forms
\mar{z48}\beq
 \lng\f^{a_1}_{p_1}\cdots \f^{a_k}_{p_k}\rng=\frac{1}{\cN}\int
\f^{a_1}_{p_1}\cdots \f^{a_k}_{p_k} \exp\{-\int
\cL(\f^a_p)d^nx\}\op\prod_p [d\phi_p^a], \label{z48}
\eeq
and the forms
\mar{z49}\ben
&& \lng\f^{a_1}_{x_1}\cdots \f^{a_k}_{x_k}\rng=\frac{1}{\cN}\int
\f^{a_1}_{x_1}\cdots \f^{a_k}_{x_k} \exp\{-\int
\cL(\f^a_{x\La})d^nx\}\op\prod_x [d\phi_x^a], \label{z49}\\
&& \cN=\int \exp\{-\int \cL(\f^a_{x\La})d^nx\}\op\prod_x
[d\phi_x^a], \nonumber\\
&& \cL(\f^a_{x\La})=\cL(x,\g_x(s^a_\La)), \nonumber
\een
which provide Euclidean Green functions. As was mentioned above,
the term $\op\prod_p [d\phi_p^a]$ in the formulas (\ref{z31}) --
(\ref{z31'}) fail to be a true measure on $T(\Bbb R^n)$ because
the Lebesgue measure on infinite-dimensional vector spaces need
not exist. Nevertheless, treated as generalization of Berezin's
finite-dimensional integrals \cite{ber}, the functional integrals
(\ref{z48}) and (\ref{z49}) restart Euclidean Green functions in
the Feynman diagram technique. Certainly, these Green functions
are singular, unless regularization and renormalization techniques
are involved.

Since a graded derivation $\vt$ (\ref{0672}) of the algebra
$\cP^0$ is a $C^\infty(X)$-linear morphism over $\id X$, it
induces the graded derivation
\mar{z50}\beq
\wh \vt_x= \g_x\circ \wh \vt\circ \g^{-1}_x: \f^a_{x\La}\to (x,
y^a_\La))\to \wh \vt^a_\La(x,y^b_\Si)\to \wh
\vt^a_\La(x,\g_x(y^b_\Si))=\wh \vt^a_{x\La}(\f^b_{x\Si})
\label{z50}
\eeq
of the range $\g_x(\cP^0)\subset B_\Phi$ of the homomorphism
$\g_x$ (\ref{z45}) for each $x\in \Bbb R^n$. The maps $\wh \vt_x$
(\ref{z50}) yield the maps
\be
&& \wh \vt_p: \f^a_p=\int \f^a_x e^{ipx}d^nx \to \int \wh
\vt_x(\f^a_x)e^{ipx}d^nx= \int \wh
\vt^a_x(\f^b_{x\Si})e^{ipx}d^nx = \\
&& \qquad  \int \wh \vt_x^a(\int (-i)^kp'_{\si_1}\cdots
p'_{\si_k}\f^b_{p'}e^{-ip'x}d_np') e^{ipx}d^nx= \wh \vt^a_p,
\qquad p\in\Bbb R_n,
\ee
and, as a consequence, the graded derivation
\be
\wh \vt(\f)=\int\f_a(p)\wh \vt(\f^a_p) d_np= \int\f_a(p)\wh
\vt^a_pd_np
\ee
of the algebra $B_\Phi$. It can be written in the symbolic form
\mar{z53,4}\ben
&& \wh \vt= \int u^a_p\frac{\dr}{\dr \f^a_p}d_np, \qquad \frac{\dr
\f^b_{p'}}{\dr \f^a_p} =\dl^b_a\dl(p'-p), \label{z53}\\
&& \wh \vt= \int u^a_x\frac{\dr}{\dr \f^a_x}d^nx, \qquad \frac{\dr
\f^b_{x'\La}}{\dr \f^a_x}= \dl^b_a\frac{\dr}{\dr x'^{\la_1}}
\cdots \frac{\dr}{\dr x'^{\la_k}} \dl(x'-x). \label{z54}
\een

Let $\al$ be an odd element, and let us consider the automorphism
\mar{bb20}\beq
\wh U=\exp\{\al \wh \vt\}=\id +\al\wh \vt \label{bb20}
\eeq
of the algebra $B_\Phi$ which can provide a change of variables
depending on $\al$ as a parameter in the functional integrals
(\ref{z48}) and (\ref{z49}) \cite{ber}. This automorphism yields a
new state $\lng.\rng'$ of $B_\Phi$ given by the equalities
\mar{bb10,'}\ben
&& \lng \f^{a_1}_{x_1}\cdots \f^{a_k}_{x_k}\rng= \lng \wh
U(\f^{a_1}_{x_1})\cdots \wh U(\f^{a_k}_{x_k})\rng'=  \label{bb10}\\
&& \qquad \frac{1}{\cN'}\int \wh U(\f^{a_1}_{x_1})\cdots \wh
U(\f^{a_k}_{x_k}) \exp\{-\int \cL(\wh
U(\f^a_{x\La}))d^nx\}\op\prod_x [d\wh
U(\phi_x^a)], \nonumber \\
&& \cN'=\int \exp\{-\int \cL(\wh U(\f^a_{x\La}))d^nx\}\op\prod_x
[d\wh U(\phi_x^a)], \nonumber\\
 && \lng \f_1\cdots \f_k\rng= \lng \wh U(\f_1)\cdots \wh
U(\f_k)\rng'= \label{bb10'}\\
&& \qquad \frac{1}{\cN'}\int \wh U(\f_1)\cdots \wh U(\f_k)
\exp\{-\int \cL_{GF}(\wh U(\f^a_p))d^nx\}\op\prod_p [d\wh
U(\phi_p^a)], \nonumber\\
&& \cN'=\int \exp\{-\int \cL_{GF}(\wh U(\f^a_p))d^nx\}\op\prod_p
[d\wh U(\phi_p^a)].\nonumber
\een
Let us apply these relations to the Green functions (\ref{z48})
and (\ref{z49}).

It follows from the decomposition (\ref{g107'}) that
\be
\int \cL(\wh U(\f^a_{x\La}))d^nx =\int (\cL(\f^a_{x\La}) + \al \wh
\vt_x^a\cE_{xa})d^nx,
\ee
where $\cE_{xa}=\g_x(\cE_a)$ are the variational derivatives. It
is a property of symbolic functional integrals that
\mar{bb21}\beq
\op\prod_x[d\wh U(\phi_x^a)]=(1+\al\int \frac{\dr \wh \vt^a_x}{\dr
\f^a_x}d^nx)\op\prod_x[d\phi_x^a]=(1+\al {\rm Sp}(\wh \vt))
\op\prod_x[d\phi_x^a]. \label{bb21}
\eeq
Then the equalities (\ref{bb10}) -- (\ref{bb10'}) result in the
identities
\mar{z62'}\ben
&& \lng\wh \vt(\f^{a_1}_{x_1}\cdots \f^{a_k}_{x_k})\rng +
\lng\f^{a_1}_{x_1}\cdots \f^{a_k}_{x_k}({\rm Sp}(\wh \vt) -\int
\wh
\vt_x^a\cE_{xa}d^nx)\rng - \label{z62'}\\
&& \qquad \lng\f^{a_1}_{x_1}\cdots \f^{a_k}_{x_k}\rng\lng {\rm
Sp}(\wh \vt) -\int \wh \vt_x^a\cE_{xa}d^nx\rng =0 \nonumber
\een
for complete Euclidean Green functions (\ref{z49}) and the similar
identities for the Green functions $\lng\f^{a_1}_{p_1}\cdots
\f^{a_k}_{p_k}\rng$ (\ref{z48}).

If $\vt$ is a variational symmetry of $L$, the identities
(\ref{z62'}) are the Ward identities
\mar{z61,2}\ben
&& \lng\wh \vt(\f^{a_1}_{p_1}\cdots \f^{a_k}_{p_k})\rng
+\lng\f^{a_1}_{p_1}\cdots \f^{a_k}_{p_k}{\rm Sp}(\wh \vt)\rng-
\lng \f^{a_1}_{p_1}\cdots \f^{a_k}_{p_k}\rng\lng{\rm Sp}(\wh
\vt)\rng =0,
\label{z61}\\
&& \op\sum_{i=1}^k(-1)^{[a_1]+\cdots +[a_{i-1}]}
\lng\f^{a_1}_{p_1}\cdots \f^{a_{i-1}}_{p_{i-1}}\wh \vt^{a_i}_{p_i}
\f^{a_{i+1}}_{p_{i+1}}\cdots \f^{a_k}_{p_k}\rng +\nonumber\\
&& \qquad \lng\f^{a_1}_{p_1}\cdots \f^{a_k}_{p_k}\int \frac{\dr
\wh \vt^a_p}{\dr \f^a_p}d_np\rng - \lng\f^{a_1}_{p_1}\cdots
\f^{a_k}_{p_k}\rng\lng\int \frac{\dr \wh \vt^a_p}{\dr
\f^a_p}d_np\rng=0,
\nonumber\\
&& \lng\wh \vt(\f^{a_1}_{x_1}\cdots \f^{a_k}_{x_k})\rng
+\lng\f^{a_1}_{x_1}\cdots \f^{a_k}_{x_k}{\rm Sp}(\wh \vt)\rng-
\lng \f^{a_1}_{x_1}\cdots \f^{a_k}_{x_k}\rng\lng{\rm Sp}(\wh
\vt)\rng =0,
\label{z62}\\
&& \op\sum_{i=1}^k(-1)^{[a_1]+\cdots +[a_{i-1}]}
\lng\f^{a_1}_{x_1}\cdots \f^{a_{i-1}}_{x_{i-1}}\wh \vt^{a_i}_{x_i}
\f^{a_{i+1}}_{x_{i+1}}\cdots \f^{a_k}_{x_k}\rng +\nonumber\\
&&\qquad \lng\f^{a_1}_{x_1}\cdots \f^{a_k}_{x_k}\int \frac{\dr \wh
\vt^a_x}{\dr \f^a_x}d^nx\rng - \lng\f^{a_1}_{x_1}\cdots
\f^{a_k}_{x_k}\rng\lng\int \frac{\dr \wh \vt^a_x}{\dr
\f^a_x}d^nx\rng=0, \nonumber
\een
generalizing the Ward (Slavnov--Taylor) identities in gauge theory
\cite{bran98,fust,grig,mccloud}. A glance at the expressions
(\ref{z61}) -- (\ref{z62}) shows that these Ward identities
generally contain anomaly because the measure terms of symbolic
functional integrals need not be $\wh\vt$-invariant. If Sp$(\wh
\vt)$ is either a finite or infinite number, the Ward identities
\mar{z63,4}\ben
&& \lng\wh \vt(\f^{a_1}_{p_1}\cdots \f^{a_k}_{p_k})\rng =
\op\sum_{i=1}^k(-1)^{[a_1]+\cdots +[a_{i-1}]}
\lng\f^{a_1}_{p_1}\cdots \f^{a_{i-1}}_{p_{i-1}}\wh \vt^{a_i}_{p_i}
\f^{a_{i+1}}_{p_{i+1}}\cdots \f^{a_k}_{p_k}\rng=0, \label{z63}\\
&& \lng\wh \vt(\f^{a_1}_{x_1}\cdots \f^{a_k}_{x_k})\rng=
\op\sum_{i=1}^k(-1)^{[a_1]+\cdots +[a_{i-1}]}
\lng\f^{a_1}_{x_1}\cdots \f^{a_{i-1}}_{x_{i-1}}\wh \vt^{a_i}_{x_i}
\f^{a_{i+1}}_{x_{i+1}}\cdots \f^{a_k}_{x_k}\rng =0 \label{z64}
\een
are free of this anomaly.

If $\vt=c^a\dr_a$, $c^a=$const, the identities (\ref{z62'}) take
the form
\mar{bb13}\ben
&& \op\sum_{r=1}^k (-1)^{[a]([a_1]+\cdots+[a_{r-1}])}
\lng\f^{a_1}_{x_1}\cdots \f^{a_{r-1}}_{x_{r-1}} \dl^{a_r}_a
\f^{a_{r+1}}_{x_{r+1}}\cdots
\f^{a_k}_{x_k})\rng -  \label{bb13}\\
&& \qquad \lng\f^{a_1}_{x_1}\cdots \f^{a_k}_{x_k}(\int
\cE_{xa}d^nx)\rng + \lng\f^{a_1}_{x_1}\cdots
\f^{a_k}_{x_k}\rng\lng \int \wh \cE_{xa}d^nx\rng =0. \nonumber
\een
One can think of them as being equations for complete Euclidean
Green functions, but they are not an Euclidean variant of he
well-known Schwinger--Dyson equations \cite{bog2}. For instance,
they identically hold if a Lagrangian $L$ is quadratic.

Clearly, the expressions (\ref{z62'}), (\ref{bb13}) are singular,
unless one follows regularization and renormalization procedures,
which however can induce additional anomaly terms.

\bigskip
\bigskip

\centerline{\bf Appendix A. Classical field theory in jet
formalism}

\bigskip

We consider a Lagrangian field system on $X=\Bbb R^n$, coordinated
by $(x^\la)$. Such a Lagrangian system is algebraically described
in terms of the following differential graded algebra (henceforth
GDA) $\cP^*$ \cite{barn,jmp05,cmp04}.

Let $Y=Y_0\oplus Y_1\to X$ be a graded vector bundle coordinated
by $(x^\la, y^a)$. Finite order jet manifolds $J^rY$,
$r=1,\ldots,$ of $Y\to X$ are also vector bundles over $X$
coordinated by $(x^\la,y^a, y^a_\La)$, $|\La|=k\leq r$, where
$\La=(\la_1,\ldots,\La_k)$ denote symmetric multi-indices. The
index $r=0$ conventionally stands for $Y$. For each $r=0,\ldots,$
we consider a graded manifold $(X,\cA_{J^rY_1})$, whose body is
$X$ and the algebra of graded functions consists of sections of
the exterior bundle
\be
\w (J^rY_1)^*=\Bbb R\op\oplus_X (J^rY_1)^*\op\oplus_X\op\w^2
(J^rY_1)^*\op\oplus_X\cdots,
\ee
where $(J^rY_1)^*$ is the dual of a vector bundle $J^rY_1\to X$.
The global basis for $(X,\cA_{J^rY_1})$ is $\{x^\la,y^a_\La\}$,
$|\La|=0,\ldots,r$. Let us consider the graded commutative
$C^\infty(X)$-algebra $\cP^0$ generated by its elements $y^a_\La$,
treated as prequantum even and odd fields and their jets. The
symbol $[a]=[y^a]=[y^a_\La]$ stands for their Grassmann parity.

Let $\gd \cP^0$ be the  Lie superalgebra of graded derivations of
the $\Bbb R$-algebra $\cP^0$, i.e.,
\be
u(ff')=u(f)f'+(-1)^{[u][f]}fu(f'), \qquad f,f'\in \cP^0, \qquad
u\in \gd\cP^0.
\ee
Its elements take the form
\mar{w5}\beq
 u=u^\la\dr_\la + \op\sum_{0\leq|\La|} u_\La^a\dr^\La_a,
 \qquad u^\la, u_\La^a
 \in \cP^0. \label{w5}
\eeq
With the Lie superalgebra $\gd \cP^0$, one can construct the
minimal Chevalley--Eilenberg differential calculus
\mar{bb1}\beq
0\to \Bbb R\to \cP^0 \ar^d \cP^1\ar^d\cdots \cP^2\ar^d\cdots
\label{bb1}
\eeq
over the $\Bbb R$-algebra $\cP^0$. It is the above mentioned  DGA
$\cP^*$ with the basis $\{y^a\}$. Its elements $\si\in \cP^k$ are
graded $\cP^0$-linear $k$-forms
\be
 \si= \op\sum \si_{a_1\ldots a_r\la_{r+1}\ldots\la_k}^{\La_1\ldots
\La_r} dy_{\La_1}^{a_1}\w\cdots\w dy_{\La_r}^{a_r}\w
dx^{\la_{r+1}}\w\cdots \w dx^{\la_k}
\ee
on $\gd \cP^0$ with values in $\cP^0$. The graded exterior product
$\w$ and the graded exterior differential, obey the relations
\be
 \si\w\si' =(-1)^{|\si||\si'| +[\si][\si']}\si'\w
\si, \qquad  d(\si\w\si')= d\si\w\si' +(-1)^{|\si|}\si\w d\si',
\ee
where $|.|$ denotes the form degree. By $\cO^*X$ is denoted the
graded differential algebra of exterior forms on $X$. There is the
natural monomorphism $\cO^*X\to \cP^*$.

Given a graded derivation $u$ (\ref{w5}) of the $\Bbb R$-algebra
$\cP^0$, the interior product $u\rfloor\si$ and the Lie derivative
$\bL_u\si$, $\si\in\cP^*$, obey the relations
\be
&& u\rfloor(\si\w\si')=(u\rfloor \si)\w\si'
+(-1)^{|\si|+[\si][u]}\si\w(u\rfloor\si'), \qquad \si,\si'\in
\cP^*, \\
&& \bL_u\si=u\rfloor d\si+ d(u\rfloor\si), \qquad
\bL_u(\si\w\si')=\bL_u(\si)\w\si'
+(-1)^{[u][\si]}\si\w\bL_u(\si').
\ee

The DGA $\cP^*$ is decomposed into $\cP^0$-modules $\cP^{k,r}$ of
$k$-contact and $r$-horizontal graded forms
\be
\si=\op\sum_{0\leq|\La_i|}\si^{\La_1\ldots \La_k}_{a_1\ldots a_k
\m_1\ldots\m_r} \th^{a_1}_{\La_1}\w\cdots\w\th^{a_k}_{\La_k}\w
dx^{\m_1}\w\cdots\w dx^{\m_r}, \qquad \th^a_\La=dy^a_\La
-y^a_{\la+\La}dx^\la.
\ee
Accordingly, the graded exterior differential on $\cP^*$ falls
into the sum $d=d_V+d_H$ of the vertical and total differentials
where $d_H\si= dx^\la\w d_\la\si$. The differentials $d_H$ and
$d_V$ and the graded variational operator $\dl$ split the DGA
$\cP^*$ into the graded variational bicomplex
\cite{barn,jmp05,cmp04}. One can think of even elements
\mar{0709}\beq
L=\cL(x^\la,y^a_\La) d^nx, \qquad \dl L= dy^a\w \cE_a
d^nx=\op\sum_{0\leq|\La|} (-1)^{|\La|} dy^a\w d_\La (\dr^\La_a
L)d^nx \label{0709}
\eeq
of the differential algebra $\cP^*$ as being a graded Lagrangian
and its Euler--Lagrange operator, respectively.

A graded derivation $u$ (\ref{w5}) is called contact if the Lie
derivative $\bL_u$ preserves the ideal of contact graded forms of
the DGA $\cP^*$. Here, we restrict our consideration to vertical
contact graded derivations, vanishing on $\cO^*X$. Such a
derivation takes the form
\mar{0672}\beq
\vt=\up^a\dr_a + \op\sum_{0<|\La|} d_\La\up^a\dr_a^\La.
\label{0672}
\eeq
It is the jet prolongation of its first summand $\up=\up^a\dr_a$.
The Lie derivative $\bL_\vt L$ of a Lagrangian $L$ (\ref{0709})
along a vertical contact graded derivation $\vt$ (\ref{0672})
admits the decomposition
\mar{g107'}\beq
\bL_\vt L= \up\rfloor\dl L +d_H\si. \label{g107'}
\eeq
One says that an odd vertical contact graded derivation $\vt$
(\ref{0672}) is a variational supersymmetry of a Lagrangian $L$ if
the Lie derivative $\bL_\vt L$ is $d_H$-exact.

\bigskip
\bigskip

\centerline{\bf Appendix B. Unbounded operators}

\bigskip

Recall that by an operator in a Hilbert space $E$ is meant a
linear morphism $a$ of a dense subspace $D(a)$ of $E$ to $E$. The
$D(a)$ is called a domain of an operator $a$. One says that an
operator $b$ on $D(b)$ is an extension of an operator $a$ on
$D(a)$ if $D(a)\subset D(b)$ and $b|_{D(a)}=a$. For the sake of
brevity, we will write $a\subset b$. An operator $a$ is said to be
bounded on $D(a)$ if there exists a real number $r$ such that
$\|ae\|\leq r\|e\|,$ $e\in D(a)$. If otherwise, it is called
unbounded. Any bounded operator on a domain $D(a)$ is uniquely
extended to a bounded and continuous operator everywhere on $E$.

An operator $a$ on a domain $D(a)$ is called closed if the
condition that a sequence $\{e_i\}\subset D(a)$ converges to $e\in
E$ and that the sequence $\{ae_i\}$ does to $e'\in E$ implies that
$e\in D(a)$ and $e'=ae$. An operator $a$ on a domain $D(a)$ is
called closable if it can be extended to a closed operator. The
closure of a closable operator $a$ is defined as the minimal
closed extension of $a$.

Operators $a$ and $b$ in $E$ are called adjoint if $\lng
ae|e'\rng=\lng e|be'\rng$, $e\in D(a)$, $e'\in D(b)$. Any operator
$a$ has a maximal adjoint operator  $a^*$, which is closed. An
operator $a$ is called symmetric if it is adjoint to itself, i.e.,
$a\subset a^*$. Hence, a symmetric operator is closable. At the
same time, the maximal adjoint operator $a^*$ of a symmetric
operator $a$ need not be symmetric. A symmetric operator $a$ is
called self-adjoint if $a=a^*$, and it is called essentially
self-adjoint if $\ol a=a^*=\ol a^*$ (ee here follow the
terminology of \cite{pow1}). For bounded operators, the notions of
symmetric, self-adjoint and essentially self-adjoint operators
coincide.

Let $E$ be a Hilbert space. The pair $(B,D)$ of a dense subspace
$D$ of $E$ and a unital subalgebra $B\subset B(E)$ of (unbounded)
operators in $E$ is called the $Op^*$-algebra ($O^*$-algebra in
the terminology of \cite{schmu}) on the domain $D$ if, whenever
$b\in B$, we have: (i) $D(b)=D$ and $bD\subset D$, (ii) $D\subset
D(b^*)$, (iii) $b^*|_D\subset B$ \cite{hor,pow1}. The algebra $B$
is provided with the involution $b\mapsto b^+=b^*|_D$, and its
elements are closable. It is important that one can associate to
an $Op^*$-algebra the von Neumann algebra which is the weak
bicommutant $B''$, where $B'=\{T\in B(E)\,|\, \lng ae,T^*e'\rng,
\, a\in B, \, e,e'\in D\}$.

A representation $\pi(A)$ of an involutive algebra $A$ in a
Hilbert space $E$ is an $Op^*$- algebra if there exists a dense
subspace $D(\pi)\subset E$ such that $D(\pi)=D(\pi(a))$ for all
$a\in A$. If a representation $\pi$ is Hermitian, i.e.,
$\pi(a^*)\subset \pi(a)^*$ for all $a\in A$, then $\pi(A)$ is an
$Op^*$-algebra. In this case, one also considers the
representations
\be
&& \ol\pi: a \to \ol\pi(a):=\ol{\pi(a)}|_{D(\ol\pi)},
\qquad D(\ol\pi)=\op\bigcap_{a\in A} D(\ol{\pi(a)}),\\
&&  \pi^*: a \to \pi^*(a):=\pi(a^*)^*|_{D(\pi^*)},
\qquad D(\pi^*)=\op\bigcap_{a\in A} D(\pi(a)^*),\\
\ee
called the closure  of a representation $\pi$ and an  adjoint
representation, respectively. There are the representation
extensions $\pi\subset\ol\pi\subset\pi^*$, where $\pi_1\subset
\pi_2$ means $D(\pi_1)\subset D(\pi_2)$. The representation
$\ol\pi$ is Hermitian, while $\pi^*=\ol\pi^*$. A Hermitian
representation $\pi(A)$ is said to be closed if $\pi=\ol\pi$, and
it is self-adjoint if $\pi=\pi^*$. Herewith, a representation
$\pi(A)$ is closed (resp. self-adjoint) if one of operators of
$\pi(A)$ is closed (resp. self-adjoint).

The representation domain $D(\pi)$ is endowed with the
graph-topology. It is generated by the neighborhoods of the origin
\be
U(M,\ve)=\{x\in D(\pi)\,:\,\op\sum_{a\in M} \|\pi(a)x\|<\ve\},
\ee
where $M$ is a finite subset of elements of $A$. All operators of
$\pi(A)$ are continuous with respect to this topology. Let us note
that the graph-topology is finer than the relative topology on
$D(\pi)\subset E$, unless all operators $\pi(a)$, $a\in A$, are
bounded \cite{schmu}. Let $\ol N^g$ denote the closure of a subset
$N\subset D(\pi)$ with respect to the graph-topology. An element
$\th\in D(\pi)$ is called strongly cyclic (cyclic in the
terminology of \cite{schmu}) if $D(\pi)\subset
\ol{(\pi(A)\th)}^g$.

In application to QFT, the following class of involutive algebras
should be mentioned. Let $A$ be a locally convex topological
involutive algebra whose topology is defined by a set of
multiplicative seminorms $p_\iota$ which satisfy the condition
$p_\iota(a^*a)=p_\iota(a)^2$, $a\in A$. It is called a
$b^*$-algebra. A unital $b^*$-algebra as like as a $C^*$-algebra
is regular and symmetric, i.e., any element $(\bb +a^*a)$, $a\in
A$, is invertible and, moreover, $(\bb +a^*a)^{-1}$ is bounded
\cite{all,igu}. The $b^*$-algebras are related to $C^*$-algebras
as follows. Any $b^*$-algebra is the Hausdorff projective limit of
a family of $C^*$-algebras, and {\it vice versa} \cite{igu}. In
particular, every $C^*$-algebra $A$ is a barreled $b^*$-algebra,
i.e., every absorbing balanced closed subset is a neighborhood of
the origin of $A$.

\bigskip
\bigskip

\centerline{\bf Appendix C. Nuclear spaces}

\bigskip

Physical applications of Hilbert spaces are limited by the fact
that the dual of a Hilbert space $E$ is anti-isomorphic to $E$.
The construction of a rigged Hilbert space describes the dual
pairs $(E,E')$ where $E'$ is larger than $E$ \cite{gelf64,piet}.

Let a complex vector space $E$ have a countable set of
non-degenerate Hermitian forms $\lng.|.\rng_k$, $k\in\Bbb N_+,$
such that
\be
\lng e|e\rng_1\leq \cdots\leq \lng e|e\rng_k\leq\cdots
\ee
for all $e\in E$. The family of norms
\mar{spr445}\beq
\|.\|_k=\lng.|.\rng^{1/2}_k, \qquad k\in\Bbb N_+, \label{spr445}
\eeq
yields a Hausdorff topology on $E$. The space $E$ is called a
countably Hilbert space if it is complete with respect to this
topology. For instance, every Hilbert space is a countably Hilbert
space where all Hermitian forms $\lng.|.\rng_k$ coincide. Let
$E_k$ denote the completion of $E$ with respect to the norm
$\|.\|_k$ (\ref{spr445}). There is the chain of injections
\mar{1086}\beq
E_1\supset E_2\supset \cdots E_k\supset \cdots \label{1086}
\eeq
together with the homeomorphism $E=\op\cap_k E_k$. The dual spaces
form the increasing chain
\mar{1087}\beq
E'_1\subset E'_2\subset \cdots \subset E'_k \subset \cdots,
\label{1087}
\eeq
and $E'=\op\cup_k E'_k$. The dual $E'$ of $E$ can be provided with
the weak$^*$ and strong topologies (we follow the terminology of
\cite{rob}). One can show that a countably Hilbert space is
reflexive.

Given a countably Hilbert space $E$ and $m\leq n$, let $T^n_m$ be
a prolongation of the map
\be
E_n\supset E\ni e\mapsto e\in E\subset E_m
\ee
to the continuous map of $E_n$ onto a dense subset of $E_m$. A
countably Hilbert space $E$ is called a nuclear space if, for any
$m$, there exists $n$ such that $T^m_n$ is a nuclear map,  i.e.,
\be
T^n_m(e)=\op\sum_i\la_i\lng e|e^i_n\rng_{E_n} e^i_m,
\ee
where: (i) $\{e^i_n\}$ and $\{e_m^i\}$ are bases for the Hilbert
spaces $E_n$ and $E_m$, respectively, (ii) $\la_i\geq 0$, (iii)
the series $\sum \la_i$ converges.

An important property of nuclear spaces is that they are perfect,
i.e., every bounded closed set in a nuclear space is compact. It
follows immediately that a Banach (and Hilbert) space is not
nuclear, unless it is finite-dimensional. Since a nuclear space is
perfect, it is separable, and the weak$^*$ and strong topologies
(and, consequently, all topologies of uniform convergence) on a
nuclear space $E$ and its dual $E'$ coincide.

Let $E$ be a nuclear space, provided with still another
non-degenerate Hermitian form $\lng.|.\rng$ which is separately
continuous, i.e., continuous with respect to each argument. It
follows that there exist numbers $M$ and $m$ such that $\lng
e|e\rng\leq M\|e\|_m,$ $e\in E$. Let $\wt E$ denote the completion
of $E$ with respect to this form. There are the injections
\mar{spr450}\beq
E\subset \wt E\subset E', \label{spr450}
\eeq
where $E$ is a dense subset of $\wt E$ and $\wt E$ is a dense
subset of $E'$, equipped with the weak$^*$ topology. The triple
(\ref{spr450}) is called the rigged Hilbert space. Furthermore,
bearing in mind the chain of Hilbert spaces (\ref{1086}) and that
of their duals (\ref{1087}), one can convert the triple
(\ref{spr450}) into the chain of spaces
\mar{1088}\beq
E\subset\cdots\subset E_k\subset\cdots E_1\subset\wt E\subset
E'_1\subset \cdots\subset E'_k\subset\cdots\subset E'.
\label{1088}
\eeq

\bigskip
\bigskip

\centerline{\bf Appendix D. Generalized functions}

\bigskip

By generalized functions were initially meant the Schwartz and
tempered distributions \cite{bogol,gelf64}.

We further follow the standard multi-index notation
\be
D^\al=\frac{\dr^{|\al|}}{\dr x^{\al_1}_1\cdots \dr x^{\al_n}_n},
\ee
where $x^{\al_i}$ are Cartesian coordinates on $\Bbb R^n$ and
$\al$ is an ordered $r$-tuple of non-negative integers
$(\al_1,\ldots,\al_r)$ and $|\al|=\al_1 +\cdots +\al_r$.

Let $\cD(\Bbb R^n)$ be the space of smooth complex functions on
$\Bbb R^n$ of compact support. It is provided with the topology
determined by the seminorms
\mar{1233}\beq
p_{\{\f_\al\}}(f)=\op{\rm sup}_x |\op\sum_\al \f_\al(x) D^a f(x)|,
\label{1233}
\eeq
where $\{\f_a\}$ are collections of smooth functions such that, on
any compact subset of $\Bbb R^n$, only a finite number of these
functions differ from zero. With this topology, $\cD(\Bbb R^n)$ is
a complete locally convex nuclear space. Its topological dual
$\cD'(\Bbb R^n)$ is called the space of Schwartz distributions on
$\Bbb R^n$  \cite{bogol,horv,piet}.

The space $\cD(\Bbb R^n)$ is a subspace of another nuclear space,
whose elements are smooth functions of rapid decrease. These are
complex smooth functions $\psi(x)$ on $\Bbb R^n$ such that the
quantities
\mar{spr453}\beq
\|\psi\|_{k,m}=\op\max_{|\al|\leq k} \op\sup_x(1+x^2)^m|D^\al
\psi(x)| \label{spr453}
\eeq
are finite for all $k,m\in \Bbb N$. The space $S(\Bbb R^n)$ of
these functions (called the Schwartz space)  is a nuclear space
with respect to the topology determined by seminorms
(\ref{spr453}) \cite{horv,piet}. Its dual $S'(\Bbb R^n)$ is the
space of tempered distributions. The corresponding contraction
form is written as
\be
\lng \psi,h\rng=\op\int \psi(x) h(x) d^nx, \qquad \psi\in S(\Bbb
R^n), \qquad h\in S'(\Bbb R^n).
\ee
The  space $S(\Bbb R^n)$ is provided with the non-degenerate
separately continuous Hermitian form
\mar{spr455'}\beq
\lng \psi|h\rng=\int \psi(x)\ol{h(x)}d^nx. \label{spr455'}
\eeq
The completion of $S(\Bbb R^n)$ with respect to this form is the
space $L^2_C(\Bbb R^n)$ of square integrable complex functions on
$\Bbb R^n$. We have the rigged Hilbert space
\be
S(\Bbb R^n)\subset L^2_C(\Bbb R^n) \subset S'(\Bbb R^n).
\ee

Let $\Bbb R_n$ be the dual of $\Bbb R^n$ and $(p_\al)$ coordinates
on $\Bbb R_n$. It is important that the Fourier transform
\mar{spr460,1}\ben
&& \psi^F(p)=\int \psi(x)e^{ipx}d^nx, \qquad px=p_\al x^\al,
\label{spr460}\\
&& \psi(x)=\int \psi^F(p)e^{-ipx}d_np, \qquad
d_np=(2\pi)^{-n}d^np, \label{spr461}
\een
defines an isomorphism between the spaces $S(\Bbb R^n)$ and
$S(\Bbb R_n)$. The Fourier transform of tempered (and Schwartz)
distributions $h$ is defined by the condition
\be
\int h(x)\psi(x)d^nx=\int h^F(p)\psi^F(-p)d_np,
\ee
written in the form (\ref{spr460}) -- (\ref{spr461}). It provides
an isomorphism between spaces of tempered distributions $S'(\Bbb
R^n)$ and $S'(\Bbb R_n)$.

Distributions are also elements of Sobolev spaces
\cite{adam,nash}. Given a domain $U\subset \Bbb R^n$, let
$L^p(U)$, $1\leq p <\infty$, be the vector space of all measurable
real functions on $U$ such that
\be
\op\int_U |f(x)|^pd^nx <\infty.
\ee
It is a Banach space with respect to the norm
\be
\|f\|_p=\left\{\op\int_U |f(x)|^pd^nx\right\}^{1/p}.
\ee
Of course, functions are identified in the space if they are equal
almost everywhere in $U$. Sobolev spaces are defined over an
arbitrary domain $U\subset \Bbb R$ as the following vector
subspaces of spaces $L^p(U)$. Let $C^k(U)$ be the space of
$k$-times differentiable function on $U$. Let us define the
functional
\mar{+600}\beq
\|f\|_{k,p}=\left\{\op\sum_{|\al|\leq k} \|D^\al
f\|^p_p\right\}^{1/p}, \qquad k=0,1,\ldots, \label{+600}
\eeq
for any $f\in C^k(U)$ on $U$ for which the right side of
(\ref{+600}) makes sense. The Sobolev space $\cH^{k,p}(U)$ is
defined as the completion of the set
\be
\{f\in C^k(U)\, :\, \|f\|_{k,p}<\infty\}
\ee
with respect to the norm (\ref{+600}). It is important that the
subset of smooth elements of $\cH^{k,p}(U)$ is dense in
$\cH^{k,p}(U)$. Conversely, the Sobolev imbedding theorem states
that $\cH^{k,p}(\Bbb R^n)\subset C^l(\Bbb R^n)$ if $k-n/p>l$. In
particular, $\cH^k\subset C^n(\Bbb R^n)$ if $k>n/2$. The Sobolev
space $\cH^{k,p}(U)$ is a separable Banach space. It is reflexive,
unless $p=1$. In physical applications, one usually deals with
Sobolev spaces of $p=2$. Let us denote $\cH^k=\cH^{k,2}(\Bbb
R^n)$. It is a separable Hilbert space with respect to the
Hermitian form
\be
\lng f|f'\rng =\op\sum_{0\leq|\al|\leq k} \int D^\al f\ol{D^\al
f'}d^nx.
\ee

By a Sobolev space is also meant the closure $W^{k,p}_0, (U)$ of a
set of smooth functions of compact support in $\cH^{k,p}(U)$. In
particular, $W^{k,p}_0(\Bbb R^n)=H^{k,p}(\Bbb R^n)$, but such an
isomorphism need not hold for an arbitrary open subset $U\subset
\Bbb R^n$.

The derivatives $D^\al$ are extended to elements $f$ of
$\cH^{k,p}(U)$ which are not differentiable functions as
distributional derivatives. Namely, $D^\al f$ is defined as a
locally integrable function on $U$ such that
\be
\op\int_U f D^\al f' d^nx=(-1)^{|\al|}\op\int_U D^\al f f' d^nx
\ee
for any smooth function $f'$ on $U$ of compact support. It follows
that such a derivative is a Schwartz distribution on $U$.

The notion of a Sobolev space is extended an arbitrary real $k$.
The Sobolev space $\cH^{k,p}(U)$, $k\in \Bbb R$, consists of those
functions and Schwartz distributions $f$ for which the norm
\be
\|f\|_{k,p}=\left\{\int |\wh f(\xi)(1+\xi^2)^{k/2}|^p
d\xi\right\}^{1/p},
\ee
where $\wh f$ is the Fourier transform of $f$, is finite. If $k$
is a non-negative integer, this definition is equivalent to the
above mentioned one. In particular, one can show that $H^{-k}$,
$k>0$, is the dual of $H^k$ so that the elements of $\cH^{-k}$ are
distributions. Namely, there are the inclusions
\be
&& \cD(\Bbb R^n)\subset \cS(\Bbb R^n)\subset\cdots\subset
\cH^k\subset\cdots \subset \cH^0=\\
&&\qquad L^0(\Bbb R^n)\subset \cH^{-1} \subset\cdots \subset
\cH^{-k}\subset\cdots \subset \cS'(\Bbb R^n)\subset \cD'(\Bbb
R^n).
\ee

Dealing with Schwartz and tempered distributions, one meets a
problem that their multiplication $hh'$ is not defined, unless
they are smooth functions $h,h'\in C^\infty(\Bbb R^n)\subset
\cD'(\Bbb R^n)$. For instance, let $\th\in S'(\Bbb R^n)$ be the
Heaviside function ($\th(x)=0$ if $x<0$, $\th(x)=1$ if $x>0$). We
have
\mar{a1}\beq
\th^2=\th. \label{a1}
\eeq
Then by differentiation
\mar{a2}\beq
2\th'\th=\th'. \label{a2}
\eeq
Multiplication by $\th$ results in $2\th'\th=\th'\th'$ and,
consequently, $2\th'=\th'$, where $\th'=\dl_0\in S'(\Bbb R^n)$ is
the Dirac delta-function. This absurd result is either a
consequence of the multiplication rule (\ref{a1}) or the
differentiation one (\ref{a2}). Keeping the operation of
differentiating distributions, one can not use the classical
product of functions for their multiplication as distributions.

To overcome this difficulty, one enlarge the space $\cD'(\Bbb
R^n)$ of Schwartz distributions to an algebra $\cG(\Bbb R^n)$,
which is the quotient of a subalgebra of moderate elements of the
ring of smooth functions on $\cD(\Bbb R^n)$ \cite{col,col1}.
Elements of $\cG(\Bbb R^n)$ are called nonlinear (or Colombeau)
generalized functions. Let $\bigodot$ denote the multiplication in
$\cG(\Bbb R^n)$. There is the canonical monomorphism
\mar{a4}\beq
C^\infty(\Bbb R^n)\to \cD'(\Bbb R^n)\to \cG(\Bbb R^n). \label{a4}
\eeq
Moreover, $C^\infty(\Bbb R^n)\to \cG(\Bbb R^n)$ is a monomorphism
of algebras, i.e., $ff'=f\bigodot f'$, $f,f'\in C^\infty(\Bbb
R^n)$. However, this property is not extended to continuous
functions. For instance, $x|x|\neq x\bigodot |x|$.

If a certain condition holds, one can associate to an element of
$\f\in \cG(\Bbb R^n)$ a distribution $\g(\f)\in \cD'(\Bbb R^n)$,
which is {\it sui generis} a projection of $\f$ on $\cD'(\Bbb
R^n)$. For instance, any distribution $h\in \cD'(\Bbb R^n)$ admits
an associated distribution which is $h$ itself. At the same time,
the generalized function $\dl_0\bigodot\dl_0$ has no associated
distribution. If $f$ and $f'$ are continuous functions, their
product $f\bigodot f'$ in $\cG(\Bbb R^n)$ admits an associated
distribution which is their product $ff'$ in $C^0(\Bbb R^n)$,
i.e., $\g(f\bigodot f')=ff'$. If $f\in C^\infty(\Bbb R^n)$ and
$h\in\cD'(\Bbb R^n)$, the product $f\bigodot h$ possesses an
associated distribution which the classical product $fh$ of the
distribution theory.

The canonical monomorphism (\ref{a4}) is lost when passing from
$\Bbb R^n$ to an arbitrary manifold. A version of the Colombeau
algebra on manifolds has been studied \cite{gross1} and applied to
different geometric constructions \cite{kunzing,stein}. Some
particular ($\cG^\infty$-regular and $\cR$-regular) classes of
nonlinear generalized functions are considered \cite{delc,ober}
and their different generalizations (e.g., generalized Sobolev
algebras) are suggested \cite{bern}.

\bigskip
\bigskip

\centerline{\bf Appendix E. Free scalar fields}

\bigskip

The following states of the Borchers algebra $A_{RS^4}$  describe
free real scalar fields of mass $m$ \cite{ccr}.

Let us provide the nuclear space $RS^4$ with the positive complex
bilinear form
\mar{qm802}\ben
&& (\psi|\psi')=\frac{2}{i}\int
\psi(x)D^-(x-y)\psi'(y)d^4xd^4y=\int \psi^F(-\om,-\op
p^\to)\psi'^F(\om,\op p^\to)\frac{d_3p}{\om},
\label{qm802}\\
&& D^-(x)=i(2\pi)^{-3}\int \exp[-ipx]\th(p_0)\dl(p^2-m^2)d^4p,
\nonumber
\een
where $D^-(x)$ is the negative frequency part of the Pauli--Jordan
function, $p^2$ is the Minkowski square, and
\be
\om=({\op p^\to}^2 +m^2)^{1/2}.
\ee
Since the function $\psi(x)$ is real, its Fourier transform
satisfies the equality $\psi^F(p)=\ol\psi^F(-p)$. The bilinear
form (\ref{qm802}) is degenerate because the Pauli--Jordan
function $D^-(x)$ obeys the mass shell equation
\be
(\Box +m^2)D^-(x)=0.
\ee
It takes nonzero values only at elements $\psi^F\in RS_4$ which
are not zero on the mass shell $p^2=m^2$. Therefore, let us
consider the quotient space $\g:RS^4\to RS^4/J$, where
$J=\{\psi\in RS^4\, :\, (\psi|\psi)=0\}$ is the kernel of the
square form (\ref{qm802}). The map $\g$ assigns to each element
$\psi\in RS^4$ with the Fourier transform $\psi^F(p_0,\op
p^\to)\in RS_4$ the couple of functions $(\psi^F(\om,\op
p^\to),\psi^F(-\om,\op p^\to))$. Let us equip the factor space
$RS^4/J$ with the real bilinear form
\mar{qm803}\ben
&& (\g\psi|\g\psi')_L={\rm Re}(\psi|\psi')= \label{qm803}\\
&& \qquad \frac12\int [\psi^F(-\om,-\op p^\to)\psi'^F(\om,\op
p^\to) +\psi^F(\om,-\op p^\to) \psi'^F(-\om,\op
p^\to)]\frac{d_3\op p^\to}{\om}. \nonumber
\een
Then it is decomposed into the direct sum $RS^4/J=L^+\oplus L^-$
of the subspaces
\be
L^\pm=\{\psi^F_\pm(\om,\op p^\to)=\frac12(\psi^F(\om,\op p^\to)
\pm\psi^F(-\om,\op p^\to))\},
\ee
which are mutually orthogonal with respect to the bilinear form
(\ref{qm803}). There exist continuous isometric morphisms
\be
&& \g_+:\psi^F_+(\om,\op p^\to) \mapsto q^F(\op
p^\to)=\om^{-1/2}\psi^F_+
(\om,\op p^\to),\\
&& \g_-:\psi^F_-(\om,\op p^\to) \mapsto q^F(\op
p^\to)=-i\om^{-1/2}\psi^F_- (\om,\op p^\to)
\ee
of spaces $L^+$ and $L^-$ to the nuclear space $RS^3$ endowed with
the nondegenerate separately continuous Hermitian form
\mar{qm807}\beq
\lng q|q'\rng=\int q^F(-\op p^\to)q'^F(\op p^\to)d_3p.
\label{qm807}
\eeq
It should be emphasized that the images $\g_+(L^+)$ and
$\g_-(L^-)$ in $RS^3$ are not orthogonal with respect to the
scalar form (\ref{qm807}). Combining $\g$ and $\g_\pm$, we obtain
the continuous morphisms $\tau_\pm: RS^4\to RS^3$ given by the
expressions
\be
&&
\tau_+(\psi)=\g_+(\g\psi)_+=\frac{1}{2\om^{1/2}}\int[\psi^F(\om,\op
p^\to) + \psi^F(-\om,\op p^\to)]\exp[-i\op p^\to\op x^\to]d_3p,\\
&&
\tau_-(\psi)=\g_-(\g\psi)_-=\frac{1}{2i\om^{1/2}}\int[\psi^F(\om,\op
p^\to) - \psi^F(-\om,\op p^\to)]\exp[-i\op p^\to\op x^\to]d_3p.
\ee

Now let us consider the CCR algebra
\mar{1110}\beq
\ccG(RS^3)=\{(\f(q),\pi(q),I),\,q\in RS^3\} \label{1110}
\eeq
modeled over the nuclear space $RS^3$, which is equipped with the
Hermitian form (\ref{qm807}). Using the morphisms $\tau_\pm$, let
us define the map
\be
RS^4\ni \psi \mapsto \f(\tau_+(\psi)) -\pi(\tau_-(\psi))\in
\ccG(RS^3).
\ee
With this map, any representation of the nuclear CCR algebra
$\ccG(RS^3)$ induces a state
\mar{qm805}\beq
f(\psi^1\cdots\psi^n)=\lng\f(\tau_+(\psi^1)) +\pi(\tau_-(\psi^1))]
\cdots [\f(\tau_+(\psi^n)) +\pi(\tau_-(\psi^n))]\rng \label{qm805}
\eeq
on the Borchers algebra $A_{RS^4}$ of scalar fields. Furthermore,
one can justify that the corresponding distributions $W_n$ fulfil
the mass shell equation and that the following commutation
relation holds:
\be
W_2(x,y) -W_2(y,x)=-iD(x-y),
\ee
where
\be
D(x)=i(2\pi)^{-3}\int
\exp[-ipx](\th(p_0)-\th(-p_0))\dl(p^2-m^2)d^4p,
\ee
is the Pauli--Jordan commutation function. Thus, the states
(\ref{qm805}) describe real scalar fields of mass $m$. For
instance, the Fock representation of the CCR algebra $\ccG(RS^3)$
define the state $f_{\rm F}$ (\ref{qm805}) which satisfies the
Wick theorem relations where $f_2$ is given by the Wightman
function
\mar{ccr21}\beq
W_2(x,y)=\frac{1}{i}D^-(x-y). \label{ccr21}
\eeq
Thus, the state $f_{\rm F}$ describe standard quantum free scalar
fields of mass $m$.

\bigskip
\bigskip

\centerline{\bf Appendix F. Wick rotation}

\bigskip

In order to describe the Wick rotation of Euclidean states, we
start with the basic formulas of the Fourier--Laplace (henceforth
FL) transformation \cite{bogol}. It is defined on Schwartz
distributions, but we focus on the tempered ones.

Let $\Bbb R^n_+$ and $\ol\Bbb R^n_+$ further denote the subset of
points of $\Bbb R^n$ with strictly positive Cartesian coordinates
and its closure, respectively. Let $f\in S'(\Bbb R^n)$ be a
tempered distribution and $\G(f)$ the convex subset of points
$q\in \Bbb R_n$ such that
\mar{1273}\beq
e^{-qx}f(x)\in S'(\Bbb R^n). \label{1273}
\eeq
In particular, $0\in\G(f)$. Let $\intr\G(f)$ and $\dr\G(f)$ denote
the interior and boundary of $\G(f)$, respectively. The FL
transform of a tempered distribution $f\in S'(\Bbb R^n)$ is
defined as the tempered distribution
\mar{7.2}\beq
f^{FL}(p+iq)=(e^{-qx}f(x))^F(p)=\int f(x)e^{i(p+iq)x} d^nx \in
S'(\Bbb R_n), \label{7.2}
\eeq
which is the Fourier transform of the distribution (\ref{1273})
depending on $q$ as parameters. One can think of the FL transform
(\ref{7.2}) as being the Fourier transform with respect to the
complex arguments $k=p+iq$.

If $\intr\G(f)\neq\emptyset$, the FL transform $f^{FL}(k)$ is a
holomorphic function $h(k)$ of complex arguments $k=p+iq$ on the
open tube $\Bbb R_n +i\intr\G(f)\subset \Bbb C_n$ over
$\intr\G(f)$. Moreover, for any compact subset $Q\subset
\intr\G(f)$, there exist strictly positive numbers $A$ and $m$,
depending on $Q$ and $f$, such that
\mar{7.3}\beq
|f^{FL}(p+iq)|\leq A(1+|p|)^m, \qquad p\in \Bbb R_n, \qquad q\in
Q. \label{7.3}
\eeq
The evaluation (\ref{7.3}) is equivalent to the fact that the
function $h(p+iq)$ defines a family of tempered distributions
$h_q(p)\in S'(\Bbb R_n)$ of the variables $p$ depending
continuously on parameters $q\in S$. If $0\in\intr\G(f)$, then
\be
f^{FL}(p+i0)=\op\lim_{q\to 0}f^{FL}(p+iq)
\ee
coincides with the Fourier transform $f^F(p)$ of $f$. The case of
$0\not\in\intr\G(f)$ is more intricate. Let $S$ be a convex domain
in $\Bbb R^n$ such that $0\in\dr S$, and let $h(p+iq)$ be a
holomorphic function on the tube $\Bbb R_n +iS$ which defines a
family of tempered distributions $h_q(p)\in  S'(\Bbb R_n)$,
depending on parameters $q$. One says that $h(p+iq)$ has a
generalized boundary value $h_0(p)\in S'(\Bbb R_n)$ if, for any
frustum $K^r\subset S\cup \{0\}$ of the cone $K\subset \Bbb R_n$
(i.e., $K^r=\{q\in K\,:\,|q|\leq r\}$), one has
\be
h_0(\psi(p))=\op\lim_{|q|\to 0,\,q\in K^r\setminus\{0\}}
h_q(\psi(p))
\ee
for all functions $\psi\in S(\Bbb R_n)$ of rapid decrease. Then
the following holds \cite{bogol}.

Let $f\in S'(\Bbb R^n)$, $\intr\G(f)\neq \emptyset$ and
$0\not\in\intr\G(f)$. A generalized boundary value of the FL
transform $f^{FL}(k)$ in $S'(\Bbb R_n)$ exists and coincides with
the Fourier transform $f^F(p)$ of the distribution $f$.

Let us apply this result to the following important case. The
support of a tempered distribution Ê$f$ is defined as the
complement of the maximal open subset $U$ where $f$ vanishes,
i.e., $f(\psi)=0$ for all $\psi\in S(\Bbb R^n)$ of support in $U$.
Let $f\in S'(\Bbb R^n)$ be of support in $\ol\Bbb R_+^n$. Then
$\ol\Bbb R_{n+}\subset \G(f)$, and the FL transform $f^{FL}$ is a
holomorphic function on the tube over $\Bbb R_{n+}$, while its
generalized boundary value in $S'(\Bbb R_n)$ is given by the
equality
\be
h_0(\psi(p))=\op\lim_{|q|\to 0,\,q\in\Bbb R_{n+}}
f^{FL}_q(\psi(p))=f^F(\psi(p)), \qquad \forall \psi\in S(\Bbb
R_n).
\ee
Conversely, one can restore a tempered distribution $f$ of support
in $\ol\Bbb R_+^n$ from its FL transform $h(k)=f^{FL}(k)$ even if
this function is known only on $i\Bbb R_{n+}$. Indeed, the
formulas
\mar{7.4,5}\ben
&& \wt h(\phi)=\op\int_{\Bbb R_{n+}}h(iq)\f(q)d_nq= \op\int_{\Bbb
R_{n+}}
d_nq \op\int_{\ol\Bbb R^n_+} e^{-qx}f(x)\f(q)d^nx= \label{7.4}\\
&& \qquad \op\int_{\ol\Bbb R^n_+} f(x)\wh\f(x)d^nx, \qquad \f\in
S(\Bbb
R_{n+}),\nonumber\\
&& \wh\f(x)=\op\int_{\Bbb R_{n+}} e^{-qx}\f(q)d_nq, \qquad
x\in\ol\Bbb R^n_+, \qquad \wh\f\in S(\ol\Bbb R^n_+), \label{7.5}
\een
define a linear continuous functional $\wt h(q)=h(iq)$ on the
space $S(\Bbb R_{n+})$. It is called the Laplace transform
$f^L(q)=f^{FL}(iq)$ of a tempered distribution $f$. The image of
the space $S(\Bbb R_{n+})$ with respect to the mapping
$\f(q)\mapsto \wh \f(x)$ (\ref{7.5}) is dense in $S(\ol\Bbb
R^n_+)$. Then the family of seminorms
$\|\f\|'_{k,m}=\|\wh\f\|_{k,m}$, where $\|.\|_{k,m}$ are seminorms
(\ref{spr453}) on $S(\Bbb R^n)$, determines the new coarsen
topology on $S(\Bbb R_{n+})$ such that the functional (\ref{7.4})
remains continuous with respect to this topology. Then the
following is proved \cite{bogol}.

The mappings (\ref{7.4}) and (\ref{7.5}) provide one-to-one
correspondence between the Laplace transforms $f^L(q)=f^{FL}(iq)$
of tempered distributions $f\in S'(\ol\Bbb R^n_+)$ and the
elements of $S'(\Bbb R_{n+})$ which are continuous with respect to
the coarsen topology on $S(\Bbb R_{n+})$.

With this correspondence, the above mentioned Wick rotation of
Green's functions of Euclidean quantum fields to causal forms in
the Minkowski space is described as follows.

Let us denote by $X$ the space $\Bbb R^4$ associated to the real
subspace of $\Bbb C^4$ and by $Y$ the space $\Bbb R^4$,
coordinated by $(y^0,y^{1,2,3})$ and associated to the subspace
$\wt Y$ of $\Bbb C^4$ whose points possess the coordinates
$(iy^0,y^{1,2,3})$. If $X$ is the Minkowski space, then one can
think of $Y$ as being its Euclidean partner. Since $X$ and $Y$ in
$\Bbb C^4$ have the same spatial subspace, we further omit the
dependence on spatial coordinates. Therefore, let us consider the
complex plane $\Bbb C^1=X\oplus iZ$ of the time $x$ and the
Euclidean time $z$ and the complex plane $\Bbb C_1=P\oplus iQ$ of
the associated momentum coordinates $p$ and $q$.

Let $W(q)\in S'(Q)$ be a tempered distribution such that
\mar{036}\beq
W=W_++W_-,\qquad W_+\in S'(\ol Q_+), \qquad W_-\in S'(\ol
Q_-).\label{036}
\eeq
For instance, $W(q)$ is an ordinary function at $0$. For every
test function $\psi_+\in S(X_+)$, we have
\mar{037}\ben
&&\frac{1}{2\pi}\op\int_{\ol Q_+}W(q)\wh\psi_+(q)dq
=\frac{1}{2\pi}\op\int_{\ol
Q_+}dq\op\int_{X_+}dx [W(q)\exp(-qx)\psi_+(x)]=\nonumber\\
&&\qquad\frac{1}{(2\pi)^2}\op\int_{\ol Q_+}dq\op\int_Pdp\op
\int_{X_+}dx[W(q)\psi^F_+(p)\exp(-ipx-qx)]=\nonumber\\
&&\qquad \frac{-i}{(2\pi)^2}\op\int_{\ol Q_+}dq
\op\int_Pdp[W(q)\frac{\psi^F_+(p)}{p-iq}]=\frac{1}{2\pi}
\op\int_{\ol Q_+}W(q)\psi^{FL}_+(iq)dq, \label{037}
\een
due to the fact that the FL transform $\psi^{FL}_+(p+iq)$ of the
function $\psi_+\in S(X_+)\subset S'(X_+)$ exists and that it is
holomorphic on the tube $P+iQ_+, Q_+$. Moreover,
$\psi^{FL}_+(p+i0)=\psi^F_+(p)$, and the function
$\wh\psi_+(q)=\psi^{FL}_+(iq)$ can be regarded as the Wick
rotation of the test function $\psi_+(x)$. The equality
(\ref{037}) can be brought into the form
\mar{038}\ben
&&\frac{1}{2\pi}\op\int_{\ol Q_+}W(q)\wh\psi_+(q)dq=
\op\int_{X_+}\wh W_+(x)\psi_+(x)dx, \label{038}\\
&&\wh W_+(x)=\frac{1}{2\pi}\op\int_{\ol Q_+}\exp(-qx)W(q)dq,
\qquad x\in X_+. \nonumber
\een
It associates to a distribution $W(q)\in S'(Q)$ the distribution
$\wh W_+(x)\in S'(X_+)$, continuous with respect to the coarsen
topology on $S(X_+)$.

For every test function $\psi_-\in S(X_-)$, the similar relations
\mar{039}\ben
&&\frac{1}{2\pi}\op\int_{\ol Q_-}W(q)\wh\psi_-(q)dq=
\op\int_{X_-}\wh W_-(x)\psi_-(x)dx, \label{039}\\
&&\wh W_-(x)=\frac{1}{2\pi}\op\int_{\ol Q_-}\exp(-qx)W(q)dq,
\qquad x\in X_-, \nonumber
\een
hold. Combining (\ref{038}) and (\ref{039}), we obtain
\mar{040}\beq
\frac{1}{2\pi}\op\int_QW(q)\wh\psi(q)dq=\op\int_X\wh
W(x)\psi(x)dx, \qquad \wh\psi=\wh\psi_++\wh\psi_-, \qquad
\psi=\psi_++\psi_-, \label{040}
\eeq
where $\wh W(x)$ is a linear functional on functions $\psi\in
S(X)$, which together with all derivatives vanish at $x=0$. One
can think of $\wh W(x)$ as being the Wick rotation of the
distribution (\ref{036}).

In particular, let a tempered distribution
\mar{1270}\beq
M(\f_1,\f_2)=\int W_2(x_1,x_2)\f_1(x_1)\f_2(x_2) \label{1270}
d^nx_1d^nx_2.
\eeq
be the Green's function of some positive elliptic differential
operator $\cE$, i.e.,
\be
\cE_{y_1}W_2(y_1,y_2)=\dl(y_1-y_2),
\ee
where $\dl$ is Dirac's $\dl$-function. Then the distribution $W_2$
reads
\mar{1272}\beq
W_2(y_1,y_2)=w(y_1-y_2), \label{1272}
\eeq
and we obtain the form
\be
&& F_2(\f_1\f_2)=M(\f_1,\f_2)=\int w(y_1-y_2)\f_1(y_1)\f_2(y_2)
d^4y_1 d^4y_2=\\
&& \qquad \int w(y)\f_1(y_1)\f_2(y_1-y)d^4y d^4y_1=\int
w(y)\vf(y)d^4y=
\int w^F(q)\vf^F(-q) d_4q, \\
&& y=y_1-y_2, \qquad \vf(y)=\int \f_1(y_1)\f_2(y_1-y)d^4y_1.
\ee
For instance, if $\cE_{y_1} =-\Delta_{y_1}+m^2$, where $\Delta$ is
the Laplacian, then
\mar{1271}\beq
w(y_1-y_2)=\int\frac{\exp(-iq(y_1-y_2))}{q^2+m^2}d_4q,
\label{1271}
\eeq
where $q^2$ is the Euclidean square, is the propagator of a
massive Euclidean scalar field. Let the Fourier transform $w^F$ of
the distribution $w$ (\ref{1272}) satisfy the condition
(\ref{036}). Then its Wick rotation (\ref{040}) is the functional
\be
\wh w(x)=\th(x)\op\int_{\ol Q_+}w^F(q)\exp(-qx)dq +
\th(-x)\op\int_{\ol Q_-}w^F(q)\exp(-qx)dq
\ee
on scalar fields in the Minkowski space. For instance, let $w(y)$
be the Euclidean propagator (\ref{1271}) of a massive scalar
field. Then due to the analyticity of
\be
w^F(q)=(q^2+m^2)^{-1}
\ee
on the domain $\im q\cdot \re q>0$, one can show that $\wh
w(x)=-iD^c(x)$ where $D^c(x)$ is a familiar causal Green's
function.

\newpage

\centerline{\bf Appendix G. Schwinger functions}

\bigskip

Let us show the difference between Schwinger functions in
Osterwalder--Shraded  Euclidean QFT and the Euclidean states of
the Borchers algebra.

As was mentioned above, the Wightman functions obey the spectrum
condition, which implies that the Fourier transform $w^F_n$ of the
distributions $w_n$ (\ref{1278}) is of support in the closed
forward light cone $\ol V_+$ in the momentum Minkowski space $\Bbb
R_4$. It follows that the Wightman function $w_n$ is a generalized
boundary value in $S'(\Bbb R^{4n-4})$ of the function
$(w^F_n)^{FL}$, which is the FL transform of the function $w^F_n$
with respect to variables $p^i_0$ and which is holomorphic on the
tube $(\Bbb R^4+iV_-)^{n-1}\subset \Bbb C^{4n}$. Accordingly,
$W_n(x_1,\ldots,x_n)$ is a generalized boundary value in $S'(\Bbb
R^{4n})$ of a function $W_n(z_1,\ldots,z_n)$, holomorphic on the
tube
\be
\{z_i\,:\, \im(z_{i+1}-z_i)\in V_-, \,\, \re z_i\in\Bbb R^4\}.
\ee
In accordance with the Lorentz covariance, the Wightman functions
admit an analytic continuation onto a wider domain in $\Bbb
C^{4n}$, called the extended forward tube. Furthermore, the
locality condition implies that they are symmetric on this domain.

Let $X$ and $Y$ be the Minkowski subspace and its Euclidean
partner in $\Bbb C^4$, respectively. Let us consider the subset
$\wt Y^n_{\neq}\subset \wt Y\subset \Bbb C^{4n}$ which consists of
the points $(z_1,\ldots,z_n)$ such that $z_i\neq z_j$. It belongs
to the domain of analyticity of the Wightman function
$W_n(z_1,\ldots,z_n)$, whose restriction to $\wt Y^n_{\neq}$
defines the symmetric function
\be
S_n(y_1,\ldots,y_n)=W_n(z_1,\ldots,z_n), \qquad
z_i=(iy^0_i,y^{1,2,3}_i),
\ee
on $Y^n_{\neq}$. It is called the Schwinger function. On the
domain $Y^n_<$ of points $(y_1,\ldots,y_n)$ such that
$0<y_1^0<\cdots< y_n^0$, the Schwinger function takes the form
\mar{1279}\beq
S_n(y_1,\ldots,y_n)=s_n(y_1-y_2,\ldots,y_{n-1}-y_n), \label{1279}
\eeq
where $s_n$ is an element of the space $S'(Y_-^{n-1})$ which is
continuous with respect to the coarsen topology on $S(Y_-^{n-1})$.
Consequently, by virtue of the formula (\ref{7.4}), the Schwinger
function $s_n$ (\ref{1279}) can be represented as
\mar{7.8}\ben
&& s_n(y_1-y_2,\ldots,y_{n-1}-y_n)= \label{7.8}\\
&& \qquad \int\exp[p^j_0(y^0_j-y^0_{j+1}) -i\op\sum^3_{k=1}p^j_k
(y^k_j- y^k_{j+1})]w_n^F(p^1,\ldots,p^n) d_4p^1\cdots d_4p^{n-1},
\nonumber
\een
where $w^F_n\in S'(\ol\Bbb R_{n+})$ is the Fourier transform of
the Wightman function $w_n$, seen as an element of $S'(\Bbb
R_{n+})$ of support in the subset $p^i_0\geq 0$. The formula
(\ref{7.8}) enables one to restore the Wightman functions on the
Minkowski from the Schwinger functions on the Euclidean space
\cite{sim74,zin}.

\newpage

\centerline{\bf Appendix H. Representations of nuclear Lie groups}

\bigskip

The typical construction is the following \cite{1}. Let $(Q,\m)$
be a localizable measurable space, where $\m$ is quasi-invariant
under a transformation group $G$. There is the unitary
representation
\mar{q1}\beq
G\ni g: u\mapsto T_L(g)u, \quad (T_L(g)u)(q)=
(d\m(gq)/d\m(q))^{1/2}u(g^{-1}q), \quad u\in L^2(Q,\m), \label{q1}
\eeq
of $G$ in the Hilbert space of $L^2(Q,\m)$ of quadratically
$\m$-integrable complex-valued functions on $Q$. The group $G$ is
equipped with the weakest topology such that the representation
(\ref{q1}) is strongly continuous. If $Q=G$ is a locally compact
group and $\m$ is the left Haar measure, this topology is weaker
than the original one.

The main mathematical results are concerned with integral
representations of continuous positive-definite functions on
commutative topological groups \cite{gelf64,1}. We mention the
following one. Let $Q$ be a real nuclear space and $Z$ a
continuous positive-definite function on $Q$ i.e.
\be
Z(q_i-q_j)\ol\al^i \al^j\geq 0, \qquad Z(0)=1,
\ee
for any $n$ elements $q_1,\ldots,q_n$ of  $Q$ and any $n$ complex
numbers $\al^1,\ldots,\al^n$. In accordance with the Bochner
theorem, any such a function is the the Fourier transform
\mar{q3}\beq
Z(q)=\op\int_{Q'}\exp[i\langle q,w\rangle]d\mu_Z(w), \label{q3}
\eeq
of some probability measure $\m_Z$ on the topological dual $Q'$ of
$Q$, and {\it vice versa} \cite{bochn,gelf64}. A continuous
positive-definite function $Z$ plays the role similar to a
reproducing kernel on a locally compact group \cite{ali,book05}.
One can think of $Q$ as being the group space of the Abelian Lie
group $G_Q$. We have the strongly continuous unitary
representation of $G_Q$ in $L^2(Q',\m_Z)$ by operators
\mar{q2}\beq
\wh q: u(w)\mapsto F_q(w)u(w), \qquad F_q(w)=\exp[i\langle
q,w\rangle]. \label{q2}
\eeq

The Hilbert space of the representation (\ref{q2}) is described as
follows. For every element $q$, the function $Z$ (\ref{q3})
defines the continuous function $Z_q(q') = Z(q'-q)$ of $q'\in Q$.
Let us consider finite linear combinations
\be
v=\op\sum^N_{i=1} v^iZ_{\f_i}, \qquad v^i\in {\Bbb C},
\ee
for all elements of $Q$ \cite{odzi}. They constitute a pre-Hilbert
space $Q_Z$ equipped with the Hermitian form
\be
(v\mid v')= Z(q_i-q_j)\ol v^i v^j.
\ee
This form is separating, and the corresponding completion of $Q_Z$
is a Hilbert space $H_Z$.

Let us consider the isometry
\be
\rho: G_Z\ni Z_q(q') \mapsto F_{q'-q}(w) \in L^2(Q',\m_Z),
\ee
which is extended to the Hilbert space $H_Z$. Then $\rho(H_Z)$
carries out the irreducible cyclic representation (\ref{q2}) of
the group $G_Q$. The cyclic element $\th_Z$ is represented by the
class of $\mu_Z$-equivalent functions $u(w)=1$ on $Q'$.
Nonequivalent measures $\m_Z$ and $\m_{Z'}$ imply different cyclic
elements $\th_K$ and $\th_{K'}$ and nonequivalent representations
(\ref{q2}).

If a measure $\m_Z$ on $Q'$ is quasi-invariant under translations
$w\mapsto w+w_q$, where elements $w_q\in Q\subset Q'$, by
definition, satisfy the relation $\langle w_q,q'\rangle =
(q',q)_Q$ for all $q'\in Q$. In this case, we have the
representation of these translations by the displacement operators
$T_L(q)$ (\ref{q1}).

\end{document}